\documentclass[twocolumn,showpacs]{revtex4}

\usepackage{graphicx,amsmath}

\newcommand{\figwidth}{0.7\columnwidth}
\newcommand{\eq}[1]{Eq.(\ref{#1})}
\newcommand{\fig}[1]{Fig.~\ref{#1}}
\newcommand{\avg}[1]{ {\langle #1 \rangle} }
\newcommand{\olcite}[1]{Ref.~\onlinecite{#1}}

\newcommand{\kb}{k_{\rm B}}
\newcommand{\ecr}{\epsilon_{\rm cr}}
\newcommand{\mucr}{\mu_{\rm cr}}

\begin{document}

\title{Liquid crystals in two dimensions: First-order phase transitions 
and nonuniversal critical behavior}

\author{R. L. C. Vink}
\affiliation{Institut f\"ur Theoretische Physik II, Heinrich Heine
Universit\"at D\"usseldorf, Universit\"atsstra{\ss}e 1, 40225
D\"usseldorf, Germany}

\date{\today}

\begin{abstract} Liquid crystals in two dimensions undergo a first-order 
isotropic-to-quasi-nematic transition, provided the particle interactions 
are sufficiently ``sharp and narrow''. This implies phase coexistence 
between isotropic and quasi-nematic domains, separated by interfaces. The 
corresponding line tension is determined, and shown to be very small, 
giving rise to strong interface fluctuations. When the interactions are no 
longer ``sharp and narrow'', the transition becomes continuous, with 
non-universal critical behavior obeying hyperscaling, and approximately 
resembling the two-dimensional Potts model. \end{abstract}

%% 64.70.Md Transitions in liquid crystals 
%% 05.50.+q Lattice theory and statistics (Ising, Potts, etc.)
%% 64.60.Fr Equilibrium properties near critical points, critical exponents 
%% 64.60.Cn Order-disorder transformations; statistical mechanics of model systems


\maketitle

Fluids consisting of elongated molecules (rods, needles) can form nematic 
phases. In the nematic phase, the molecules are aligned, which 
distinguishes it from the isotropic phase, where the molecular 
orientations are random. The nematic phase is the standard example of a 
liquid crystal. Consequently, its properties are of fundamental 
importance. The nematic phase can be prepared in several ways. In 
thermotropic liquid crystals, the isotropic-to-nematic (IN) transition is 
temperature driven: starting in the high temperature isotropic phase, the 
nematic phase is reached by lowering the temperature. The Lebwohl-Lasher 
(LL) model \cite{physreva.6.426} provides a convenient theoretical 
framework. In this model, a unit vector $\vec{d}_i$ is assigned to each 
site $i$ of a lattice, where $\vec{d}_i$ represents the orientation of the 
molecule at the $i$-th lattice site. The molecules interact with 
Hamiltonian
\begin{equation}\label{eq:ll}
 {\cal H} =  - \epsilon 
 \sum_{\langle i,j \rangle} | \vec{d}_i \cdot \vec{d}_j |^p,
\end{equation}
where the sum is over nearest neighbors (factors of $\kb T$ are absorbed 
in the coupling constant $\epsilon$, with $T$ the temperature, and $\kb$ 
the Boltzmann constant). Note that \eq{eq:ll} is invariant under 
inversion, $\vec{d}_i \to -\vec{d}_i$, which is characteristic of liquid 
crystals. In the original paper, \eq{eq:ll} is studied in $d=3$ dimensions 
with $p=2$ \cite{physreva.6.426}. In this case, at low enough temperature, 
\eq{eq:ll} undergoes a first-order transition from an isotropic phase to a 
nematic phase. Nematic phases also occur in lyotropic systems, where 
density drives the IN transition. In a seminal paper \cite{onsager:1949}, 
Onsager showed that infinitely slender rods in $d=3$ dimensions also 
undergo a first-order transition from an isotropic to a nematic phase, at 
sufficiently high density.

The IN transition in $d=3$ dimensions is well understood. In contrast, the 
two-dimensional case, which will be the topic of the present 
investigation, is more controversial. More precisely, we consider 
\eq{eq:ll} in $d=2$ spatial dimensions, using two-component unit vectors 
$\vec{d}_i$. For $p=2$, \eq{eq:ll} then becomes the XY model 
\cite{kosterlitz:1974}, and a nematic phase with true long-range order is 
ruled out by the Mermin-Wagner (MW) theorem \cite{physrevlett.17.1133}. 
The two-dimensional XY model and its variants were thought to be without a 
phase transition for a long time, until Kosterlitz and Thouless (KT) 
proved that a phase transition {\it does} occur, and clarified its 
topological nature \cite{kosterlitz.thouless:1973}. The KT transition is 
one from a (high temperature) isotropic phase, with exponential decay of 
the angular correlations, to a (low temperature) quasi-nematic phase, with 
power-law decay of the correlations. In the XY model, the KT transition is 
continuous. By lowering the temperature, starting in the isotropic phase, 
the correlation length grows exponentially, until it diverges at the 
transition temperature, where the quasi-nematic phase sets in. Since the 
quasi-nematic phase has infinite correlation length, it is a critical 
phase. Consequently, the order parameter $\Delta$ and the susceptibility 
$\chi$ scale as
\begin{equation}\label{eq:sc}
  \Delta \propto L^{-\beta/\nu}, \hspace{1cm} \chi \propto L^{\gamma/\nu},
\end{equation}
with $L$ the system size; $\beta$, $\gamma$, and $\nu$ are the critical 
exponents of the order parameter, susceptibility, and correlation length, 
respectively. Since the correlation length diverges exponentially, the 
exponents themselves are undefined. However, exponent {\it ratios} are 
still defined \cite{kosterlitz:1974, archambault.bramwell.ea:1997}, via 
\eq{eq:sc}.

Since the XY model and the LL model are similar, the IN transition in 
two-dimensional liquid crystals is often assumed to be of the conventional 
KT type. Clearly, for \eq{eq:ll} with $p=2$ this is justified. If one then 
accepts a well-defined universality class for two-dimensional liquid 
crystals, it seems reasonable that \eq{eq:ll} for arbitrary $p \geq 2$, 
and indeed also lyotropic liquid crystals, are all qualitatively similar. 
The purpose of this Letter is to demonstrate that the IN transition in two 
dimensions is far more subtle. Our results are inspired by {\it 
generalized} XY models in $d=2$ \cite{physrevlett.88.047203}, for which it 
has been proved that the KT transition can become first order 
\cite{physrevlett.89.285702}. Since \eq{eq:ll} can be mapped onto the 
generalized XY model, using $(\cos x)^{2p} = 2^{-p} \, (1 + \cos(2x))^p$, 
the consequences of this result should be relevant for liquid crystals as 
well. Indeed, for liquid crystals interacting via \eq{eq:ll} with large 
$p$, a first-order transition is found, including a phase coexistence 
region, characterized by a finite line tension. For smaller $p$, 
non-universal critical behavior is found, with exponent ratios that vary 
continuously with $p$, while obeying hyperscaling. Interestingly, the 
variation of the exponent ratios we observe is qualitatively similar to 
that of the $d=2$ Potts model \cite{revmodphys.54.235}. Finally, we 
consider an {\it off-lattice} liquid crystal, for which similar 
non-universal critical behavior is found.

\begin{figure}
\begin{center}
\includegraphics[clip=,width=\figwidth]{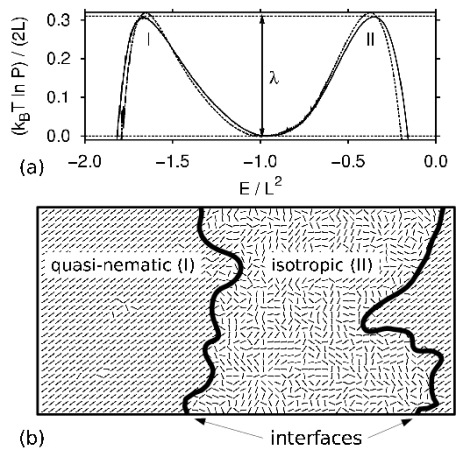} 
\caption{\label{energy} Evidence of a first-order phase transition in the 
thermotropic liquid crystal of \eq{eq:ll} with $p=1000$ at $\epsilon 
\approx 2.5$. (a) Scaled and shifted logarithm of $P(E)$ for $L=10$ (solid 
curve) and $L=15$ (dashed curve); $\lambda$ reflects the line tension of 
the IN interface. (b) Typical snapshot obtained in the coexistence 
region.}
\end{center}
\end{figure}

We begin our investigation by considering \eq{eq:ll} in the limit of large 
$p$. In this case, the nearest neighbor interaction is extremely ``sharp 
and narrow''. In other words, two neighboring molecules lower the energy 
{\it only} when they are closely aligned; otherwise, the interaction 
quickly vanishes, which mimics the Kronecker-$\delta$ Hamiltonian of the 
Potts model \cite{revmodphys.54.235}. In fact, for large $p$, \eq{eq:ll} 
qualitatively resembles the $q$-state Potts model, with $q \propto 
p^{1/2}$ \cite{physrevlett.88.047203}. The $q$-state Potts model in $d=2$ 
exhibits a first-order phase transition when $q>4$, and so one may hope to 
see a similar transition in \eq{eq:ll} when $p$ becomes large. To this 
end, we have Monte Carlo simulated \eq{eq:ll} with $p=1000$, on a 
periodic $L \times L$ lattice. The simulations are performed using 
standard single-particle Metropolis moves, combined with a biased sampling 
scheme \cite{virnau.muller:2004}, and histogram reweighting 
\cite{ferrenberg.swendsen:1989}. Evidence of a first-order transition is 
obtained from the distribution $P(E)$, defined as the probability to 
observe the energy $E$ during the simulation 
\cite{billoire.neuhaus.ea:1994}. For $\epsilon \approx 2.5$ and $p=1000$, 
we find that $P(E)$ becomes bimodal, see \fig{energy}(a). The peak at low 
energy (I) reflects the quasi-nematic phase, the peak at high energy (II) 
the isotropic phase, and the region in between the peaks corresponds to 
phase coexistence (the latter is characteristic of first-order 
transitions). Simulation snapshots obtained in the region where $P(E)$ 
attains its minimum strikingly confirm phase coexistence, see 
\fig{energy}(b). Here, a rectangular lattice was used, such that the 
interfaces form parallel to the short edge of the lattice, since this 
minimizes the total amount of interface in the system. Note that the 
coexistence is between an isotropic and a quasi-nematic phase: both phases 
lack long-range order in the thermodynamic limit. The decay of nematic 
order with system size in the quasi-nematic phase may, however, be very 
slow, giving the impression of a true nematic phase 
\cite{bramwell.holdsworth:1993}. The interfaces in \fig{energy}(b) are not 
flat, and appear to be decorated with capillary waves. Our results even 
allow for an estimate of the line tension $\lambda$ between the coexisting 
domains. To this end, note that we have plotted the logarithm of $P(E)$ 
divided by $2L$ in \fig{energy}(a), with the minimum between the peaks 
shifted to zero. The height of the peaks then reflects the line tension 
\cite{binder:1982}. The results of both system sizes are remarkably 
consistent, and yield $\lambda \approx 0.3 \, \kb T$ per lattice spacing.

Next, we investigate what happens when the interaction of \eq{eq:ll} is no 
longer ``sharp and narrow''. Clearly, for $p=2$ in \eq{eq:ll}, critical 
behavior of the XY model should be detected. Therefore, somewhere in the 
interval $2 < p < 1000$, a crossover to first-order behavior must take 
place. We have performed additional simulations of \eq{eq:ll}, and find 
that for $p<50$, bimodal energy distributions $P(E)$ can no longer be 
identified. In fact, the case $p=50$ is near the borderline: for small 
systems, bimodal energy distributions do occur, at $\epsilon \approx 
1.86$, but the free energy barrier $\Delta F \equiv 2 L \lambda$ decreases 
with $L$, and so the bimodal structure does not survive the thermodynamic 
limit. Following \olcite{physrevlett.65.137}, we conclude that the 
transition in \eq{eq:ll} is no longer first-order when $p \leq 50$. 
Interestingly, a similar phenomenon also occurs in the two-dimensional 
$q$-state Potts model \cite{revmodphys.54.235}. Here, $q=4$ is the 
borderline case: when $q \leq 4$, the Potts model no longer exhibits a 
first-order phase transition. Instead, the transition becomes critical, 
with critical exponents that depend on $q$. If we accept that $p$ in 
\eq{eq:ll} is analogous to the number of Potts states $q$, then $p=50$ 
roughly corresponds to $q=4$. It then becomes of interest to investigate 
the critical behavior of \eq{eq:ll} in the regime $p \leq 50$. If the 
analogy to the Potts model remains valid, critical exponents that depend 
on $p$ are to be expected.

\begin{figure}
\begin{center}
\includegraphics[clip=,width=\figwidth]{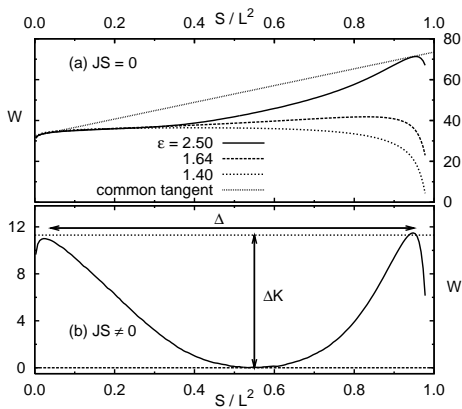}
\caption{\label{prob} Nematic order parameter distributions $W \equiv \ln 
P(S)$ of \eq{eq:ll} for $p=L=10$. (a) $W$ in the {\it absence} of $JS$, 
for several $\epsilon$, including the ``common tangent'' construction for 
$\epsilon=2.5$. (b) Bimodal form of $W$ for $\epsilon=2.5$, in the {\it 
presence} of $JS$, including the definition of $\Delta K$ and $\Delta$.}
\end{center}
\end{figure}

To this end, we now consider \eq{eq:ll} using $p=10$. Since $P(E)$ is no 
longer bimodal, a different quantity must be used to locate the phase 
transition. For liquid crystals, a natural choice is the nematic order 
parameter $S$, defined as the maximum eigenvalue of the orientational 
tensor $Q_{\alpha\beta} = \sum_{i=1}^N \left( 2 d_{i\alpha} d_{i\beta} - 
\delta_{\alpha\beta} \right)$, with $d_{i\alpha}$ the $\alpha$ component 
($\alpha = x,y$) of the orientation $\vec{d}_i$ of molecule $i$, 
$\delta_{\alpha\beta}$ the Kronecker delta, and $N$ the total number of 
particles. For disordered phases, $S$ will be small; for ordered phases 
with strong alignment, $S$ will be larger. We simulate \eq{eq:ll} as 
before, and measure the distribution $P(S)$, defined as the probability to 
observe the nematic order parameter $S$. First evidence of a phase 
transition is provided in \fig{prob}(a). Shown is $W \equiv \ln P(S)$ for 
several values of $\epsilon$. The striking feature of \fig{prob}(a) is the 
formation of a ``kink'' in $W$, when $\epsilon$ is sufficiently large. It 
then becomes possible to perform a ``common tangent'' construction. We 
thus find a rather special phase transition, characterized by a change in 
the shape of $W$. The transition is one from a high-temperature (low 
$\epsilon$) phase, where ``common tangents'' in $W$ do not occur, to a 
low-temperature (high $\epsilon$) phase, where they do. We note in passing 
that the probability distribution of the magnetization in the 
two-dimensional XY model shows similar behavior 
\cite{archambault.bramwell.ea:1997}. In other words, the formation of a 
``kink'' does not imply long-range order in \eq{eq:ll}, which indeed is 
ruled out by the MW theorem \cite{physrevlett.17.1133}.

\begin{figure}  
\begin{center}
\includegraphics[clip=,width=\figwidth]{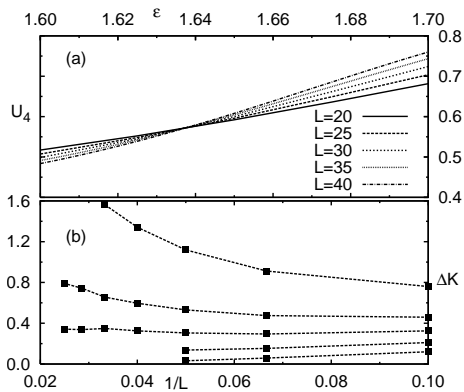}
\caption{\label{cumulant} Finite size scaling analysis of \eq{eq:ll} with 
$p=10$. (a) $U_4$ as a function of $\epsilon$ for several system sizes 
$L$. The intersection point yields an estimate of $\ecr$. (b) $\Delta K$ 
as a function of $1/L$, for $\epsilon = 1.70; 1.66; 1.64; 1.62; 1.60$ 
(top to bottom).}
\end{center}
\end{figure}

\begin{figure}
\begin{center}
\includegraphics[clip=,width=\figwidth]{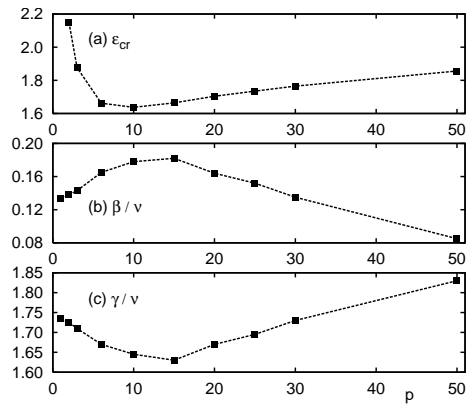}
\caption{\label{crit} Critical properties of \eq{eq:ll} as a function of $p$.}
\end{center}
\end{figure}

To accurately locate the value of $\epsilon$ above which the ``kink'' in 
$W$ begins to form, it is convenient to add a term $J S$ to the 
Hamiltonian, such that \eq{eq:ll} becomes ${\cal H} = - \epsilon 
\sum_{\langle i,j \rangle} | \vec{d}_i \cdot \vec{d}_j |^p + JS$, with $S$ 
the nematic order parameter, and coupling constant $J$. The effect of 
$J>0$ is to penalize the nematic phase. Provided a ``kink'' is present, 
$W$ can be cast into bimodal form. \fig{prob}(b) gives an example for 
$\epsilon=2.5$. Here, $J$ was tuned using $P_{J \neq 0}(S) = P_{J=0}(S) 
\exp(-J S)$, such that the two peaks in $P(S)$ were of equal area 
\cite{binder.landau:1984}. The magnitude of the ``kink'' now shows up as a 
barrier, marked $\Delta K$ in \fig{prob}(b). To locate the transition, we 
use the finite size scaling approach of \olcite{physrevlett.65.137}. To 
this end, $\Delta K$ is measured as a function of $\epsilon$ and $L$. The 
result is shown in \fig{cumulant}(b). For small $\epsilon$, $\Delta K$ 
decreases with system size, implying that the ``kink'' does not survive 
the thermodynamic limit. For large $\epsilon$, $\Delta K$ increases with 
system size, in which case the ``kink'' survives. For intermediate 
$\epsilon$, $\Delta K$ remains roughly constant, implying that the 
transition from the low-$\epsilon$ ``kink-less'' phase, to the 
high-$\epsilon$ ``kink'' phase, passes through a critical point 
\cite{physrevlett.65.137}, at $\epsilon \approx 1.64$. Further 
confirmation is obtained from the Binder cumulant $U_4 \equiv \avg{m^2}^2 
/ \avg{m^4}$ of the bimodal distributions, with $m = S - \avg{S}$, see 
\fig{cumulant}(a). At a critical point, the cumulant becomes 
$L$-independent \cite{binder:1981}. The sharp intersection point in 
\fig{cumulant}(a) provides strong evidence that \eq{eq:ll} indeed becomes 
critical. For $p=10$, criticality is obtained at $\ecr \approx 1.637$, in 
excellent agreement with the previous estimate.

Having established the critical value of $\epsilon$, critical exponent 
ratios can be measured in several ways. For example, $\beta/\nu$ is 
obtained by fitting the decay of the order parameter $\Delta$ at $\ecr$ to 
\eq{eq:sc}, with $\Delta$ defined in \fig{prob}(b). Similarly, 
$\gamma/\nu$ follows from the finite size behavior of the susceptibility 
$\chi = (\avg{S^2} - \avg{S}^2)/L^2$ at $\ecr$, see \eq{eq:sc}. In 
addition, exponent ratios can be obtained using the method of Loison 
\cite{loison:1999}. We find that both techniques are remarkably 
consistent: for $p=10$ in \eq{eq:ll}, we obtain $\beta/\nu \approx 0.175$ 
and $\gamma/\nu \approx 1.645$. Note that these ratios strikingly obey the 
hyperscaling relation $\gamma/\nu + 2 \beta/\nu = d$.  By repeating the 
above analysis for different values of $p$, the result of \fig{crit} is 
obtained. Shown are $\ecr$ (a), $\beta/\nu$ (b), and $\gamma/\nu$ (c), as 
a function of $p$. We first note that the exponent ratios for $p \neq 10$ 
also obey hyperscaling.  This result is important because it demonstrates 
the consistency of our data, and provides additional confirmation that 
\eq{eq:ll}, for small $p$, indeed becomes critical at the transition 
point. An even more striking feature is that the exponent ratios depend on 
$p$. Such non-universal critical behavior may seem surprising, but has 
been observed before in different systems \cite{physrevb.44.4819, 
physrevlett.43.737}. Indeed, based on the analogy to the Potts model, 
$p$-dependent critical behavior was already anticipated. In fact, the 
non-monotonic variation of the exponent ratios we observe in \fig{crit}, 
is also characteristic of the $q$-state Potts model. In this case, the 
exponent ratios assume their extrema at $q \approx 3.33$. Our results thus 
suggest that \eq{eq:ll} for $p \sim 10-20$, roughly corresponds to a $q 
\sim 3$ Potts model. The actual exponent values of the Potts model, 
however, do not agree with those of \fig{crit}. In other words, the 
analogy between \eq{eq:ll} and the Potts model is not exact. Note that the 
rotation symmetry of \eq{eq:ll} is {\it not} the permutation symmetry of 
the Potts model. An exact correspondence is therefore not to be expected. 
This is also manifested by the behavior of $\ecr$, see \fig{crit}(a). 
Whereas the increase of $\ecr$ with $p$ for large $p$ is consistent with 
the Potts model, the decrease at small $p$ is not. In fact, for $p=2$ in 
\eq{eq:ll}, we expect XY critical behavior to occur. As \fig{crit}(c) 
demonstrates, $\gamma/\nu$ is indeed close to the XY value $7/4$ 
\cite{kosterlitz:1974} in that case.

We now consider an off-lattice liquid crystal, namely a fluid of soft 
rods. The rods are defined as rectangles, of length $l$ and width $w$, 
capped at each end by a semi-circle of diameter $w$ (we set $l/w=16$, and 
$l$ will be the unit of length). The rods interact via a repulsive pair 
potential, whereby rod overlap is penalized with energy cost $\kappa=2$. 
The rods are simulated in the grand-canonical ensemble, i.e.~at constant 
temperature $T$, chemical potential $\mu$, and system area $A$, while the 
number of particles $N$ fluctuates. We use a simulation square of size $L 
\times L$, with periodic boundary conditions. While for \eq{eq:ll} a phase 
transition occurs above a certain $\epsilon$, here that role is played by 
$\mu$. This also implies that the analogue of energy in \eq{eq:ll}, 
becomes the number of particles $N$, since $\mu$ couples to $N$ in the 
grand-canonical ensemble. The probability distributions $P(S)$ and $P(N)$ 
were measured for various $\mu$. We find that bimodal distributions $P(N)$ 
do {\it not} occur, strongly suggesting that a first-order transition is 
absent. In other words, soft rods resemble \eq{eq:ll} in the limit of 
small $p$. Indeed, we find that $\ln P(S)$ develops a ``kink'' when $\mu$ 
exceeds a critical value $\mucr$. For $\mu > \mucr$, bimodal distributions 
$\ln P(S)$ can be realized, as in \fig{prob}(b), by tuning $J$. Finite 
size scaling confirms that the system becomes critical at the point where 
the ``kink'' first appears. For soft rods, $\mucr \approx 1.985$, 
$\beta/\nu \approx 0.17$, and $\gamma/\nu \approx 1.65$ are obtained, 
consistent with hyperscaling. By comparing to \fig{crit}, we conclude that 
soft rods resemble \eq{eq:ll} with $p \sim 10-20$.

In summary, for a {\it lattice} liquid crystal with ``sharp and narrow'' 
interactions, the existence of a first-order transition was shown, 
including an estimate of the line tension between the coexisting domains. 
The line tension is small, giving rise to strong interface fluctuations. 
When the interaction is no longer ``sharp and narrow'', the transition 
becomes continuous. Liquid crystals then show non-universal critical 
behavior, with exponent ratios that depend on the ``sharpness'' of the 
interaction, but that do obey hyperscaling. In addition, the behavior of 
the exponent ratios follows the Potts trend. For {\it lattice} liquid 
crystals, the transition type can be selected using the parameter $p$ in 
\eq{eq:ll}. For {\it off-lattice} systems, such a parameter is not so 
easily identified. In soft rods, the interaction is clearly not ``sharp 
and narrow'' enough to induce a first-order transition. It remains of 
interest to identify {\it off-lattice} interactions that do facilitate 
first-order transitions in two-dimensional liquid crystals. Possibly, this 
could be achieved using Gay-Berne type potentials \cite{gay:3316}. A 
different application could be to use the present simulation methodology 
to study melting in two dimensions. Such investigations are a topic for 
future work.

\acknowledgments

This work was supported by the {\it Deutsche Forschungsgemeinschaft} under 
the SFB-TR6/D3. I thank H.~Wensink, H.~L\"owen, A.~van Enter, K.~Binder, 
and D.~Frenkel for stimulating discussions and/or helpful suggestions.

\bibstyle{revtex}
\bibliography{mainz}

\end{document}